\documentclass{KapProc} 
\kluwerbib
\let\lcitebracket(
\let\rcitebracket)
\usepackage[dvips]{graphics}
\newcommand{\lsim}{\raise0.3ex\hbox{$<$}\kern-0.75em{\lower0.65ex\hbox{$\sim$}}}
\newcommand{\gsim}{\raise0.3ex\hbox{$>$}\kern-0.75em{\lower0.65ex\hbox{$\sim$}}}
\begin{document}


\articletitle{QSO environments at intermediate redshifts}

\author{Margrethe Wold$^1$, Mark Lacy$^2$, Per B. Lilje$^3$, S. Serjeant$^{4,5}$} 
\affil{$^1$Stockholm Observatory, University of Stockholm, Sweden\\
$^2$IGPP, Lawrence Livermore National Laboratory and UC Davis, California\\
$^3$Institute of Theoretical Astrophysics, University of Oslo\\
$^4$Astrophysics Group, Imperial College London\\
$^5$Unit for Space Sciences and Astrophysics, University of Kent at Canterbury}
\begin{abstract}
\noindent
We have made a survey of quasar environments at $0.5 \leq z \leq 0.8$, 
using a sample of both radio-loud and radio-quiet quasars 
matched in $B$-band luminosity. Our observations include images of 
background control fields to provide a good determination of the
field galaxy counts. About 10 per cent of the quasars appear to 
live in rich clusters, whereas approximately 45 per cent live in 
environments similar to that of field galaxies. 

The richness of galaxies within a 0.5 Mpc radius around the radio-quiet quasars
is found to be indistinguishable from the richness around the radio-loud quasars, 
corresponding on average to groups or poorer clusters of galaxies. 
Comparing the galaxy richness in the radio-loud quasar fields with 
quasar fields in the literature, we find no evidence of an evolution in 
the environment with epoch. Instead, a weak, but significant correlation 
between quasar radio luminosity and environmental richness is present. 
It is thus possible that the environments of quasars, at least the 
powerful ones, do not evolve much between the present epoch and $z\approx0.8$.

\end{abstract}


\section*{Sample selection and observations}

\begin{figure}[h]
\includegraphics{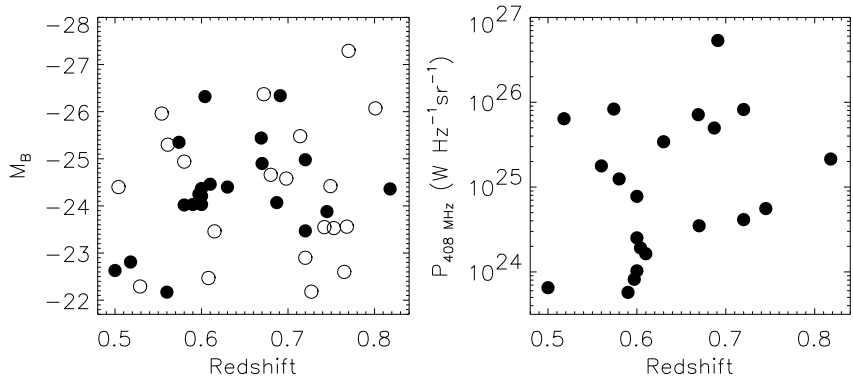}
\caption{Left: quasar absolute magnitudes as a function of redshift. 
Open circles correspond to RQQs, and filled circles correspond 
to RLQs. Right: radio luminosity as a function of redshift for 
the RLQs.}
\label{figure:fig1}
\end{figure}

We report on a survey of the galaxy environments
of intermediate redshift radio-loud and radio-quiet quasars
(RLQs and RQQs, hereafter), 
described in more detail by Wold et al.\ (2000a; 2000b). 
The quasar sample was pulled from complete catalogues with different flux
limits, in order to cancel the correlation between redshift
and luminosity known to exist in flux-limited surveys. It contains 
20 RQQs and 21 steep-spectrum RLQs within a narrow redshift range 
($0.5 \leq z \leq 0.8$), but with a wide range in AGN luminosity, 
both optical and radio. The two quasar samples are matched in 
both redshift and optical luminosity, and the distribution of 
sources in the redshift--optical/radio luminosity
plane is shown in Fig.~\ref{figure:fig1}. Our assumed cosmology has 
$H_{0}=50$ km\,s$^{-1}$\,Mpc$^{-1}$, $\Omega_{0}=1$ and $\Lambda=0$. 

The RLQs were selected from two complete surveys limited
both in the optical and in the radio; low radio luminosity quasars from 
the 7CQ survey \cite{riley99} and high radio luminosity sources from 
the MRC-APM survey (Serjeant 1996; Maddox et al.\ in prep.; Serjeant 
et al.\ in prep.)
The RQQs were selected from three different optically 
selected surveys, the faint UVX survey by Boyle et al.\ (1990), 
the intermediate-luminosity LBQS \cite{hfc95} and the 
bright BQS \cite{sg83}. 

Most of the quasar fields were imaged with the 2.56-m Nordic Optical Telescope
on LaPalma, Spain, but there are also some data from the HST and the 107-in 
telescope at the McDonald Observatory. In addition to the quasar
fields, we also imaged several background control fields by offsetting the 
telescope 5--10 arcmin away from the quasar targets. 
Our observing strategy was to use filters that target
emission longwards of the rest-frame 4000 {\AA} break at the quasar
redshifts, and thereby give preference to galaxies with 
evolved stellar populations physically associated with the quasars. In some
cases we also used two filters in order to straddle the redshifted 4000 {\AA}
break. Depending on whether the quasar had a redshift $z<0.67$
or $z \geq 0.67$, we used $V$ and $R$, or $R$ and $I$, respectively.

Using the galaxy counts from the background control fields, we
evaluated the number of excess galaxies 
within a radius of 0.5 Mpc centered on the quasar, and 
thereafter converted to the `clustering amplitude', $B_{\rm gq}$,
the amplitude of the spatial galaxy--quasar cross correlation function.
The analysis involving the clustering amplitude is
discussed elsewhere (e.g.\ Longair \& Seldner 1979; Yee \& Green 1987; 
Yee \& L{\'o}pez-Cruz 1999).

\begin{figure}[h]
\includegraphics{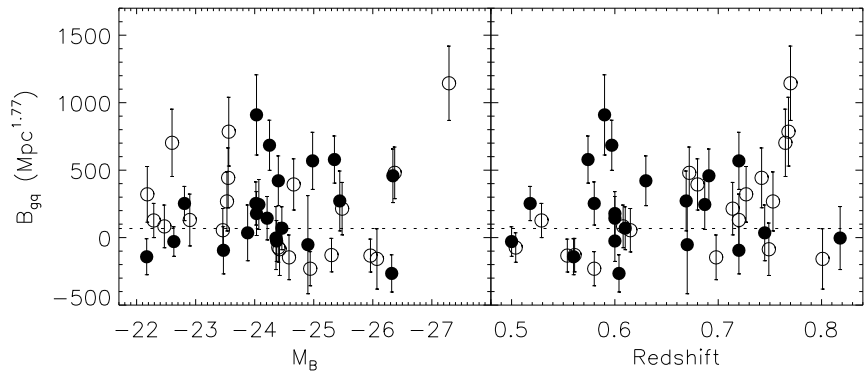}
\caption{Clustering amplitudes as a function of quasar
absolute magnitude (left) and redshift (right). The open circles
correspond to the RQQs, and the filled circles
are the RLQs. The dotted line indicates the
clustering amplitude for local field galaxies (Davis \& Peebles 1983).}
\label{figure:fig2}
\end{figure}

\section*{Radio-loud vs radio-quiet quasars}

The clustering amplitude for each quasar field is plotted 
as a function of quasar absolute $B$ magnitude and
redshift in Fig.~\ref{figure:fig2}. As seen from the figure, there is
a wide spread in the amplitudes. About 10 per cent of the quasars
seem to live in very rich environments ($B_{\rm gq} \gsim 700$ Mpc$^{1.77}$), in some cases
perhaps corresponding to Abell class 1--2 clusters. 
Another 10 per cent live in fairly rich environments, with $B_{\rm gq}$ between 500 and 700, 
and 45 per cent in environments similar to the field,
$B_{\rm gq} \lsim 100$ Mpc$^{1.77}$.

Interestingly, we find that the average environment of the RQQs is 
indistinguishable from that of the RLQs. The mean clustering amplitudes for the
RLQ and the RQQ samples are 210$\pm$82 and 213$\pm$66 Mpc$^{1.77}$, respectively. 
We thus find that the mean clustering amplitudes for the
RLQ and the RQQ samples are statistically
indistinguishable, implying that the RLQs and the RQQs
live in similar environments at these redshifts. 
This result disagrees with Ellingson, Yee \& Green (1991) who found that
at $0.3 < z < 0.6$, the RLQs exist more often in rich
environments than the RQQs, perhaps due to subtle selection effects
in their somewhat heterogeneous sample. But our result is consistent with recent host galaxy
studies finding that powerful RQQs exist in luminous, massive elliptical
galaxies similar to the RLQs (e.g.\ Bahcall et al.\ 1997; 
McLure et al.\ 1999). Other investigators are also finding
similar environments 
for powerful RLQs and RQQs at $z\approx0.2$ (Fisher et al.\ 1996; 
McLure \& Dunlop 2000; these proceedings). 

\begin{figure}[h]
\includegraphics{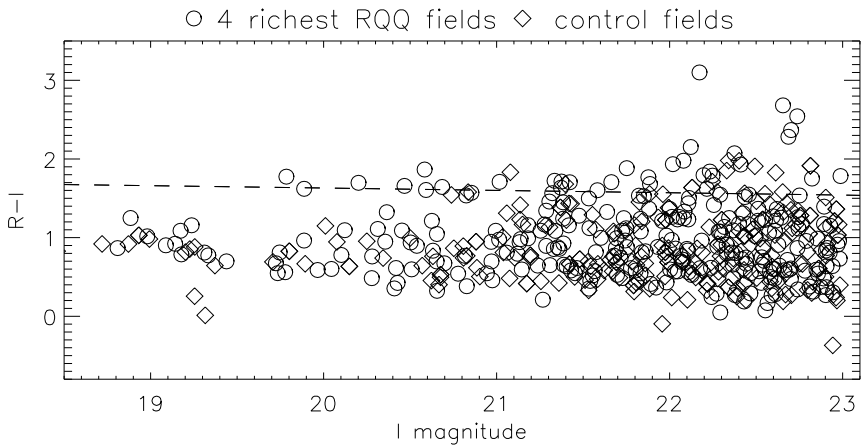}
\caption{Colour-magnitude diagram of galaxies in the four richest RQQ fields (circles)
and in the background control fields (diamonds). There is a hint of a red sequence 
(dashed line) in the quasar fields which is not present in the background control fields. 
The mean redshift of the four RQQs is 0.74, and the colour of the red sequence 
corresponds to the expected colour of galaxies at $z\approx0.7$--0.8.}
\label{figure:fig3}
\end{figure}

Note that both the quasar fields and the background
comparison fields were obtained in exactly the same manner, and that
we have applied the same analysis to both samples. The comparison
between the RLQ and the RQQ environment is therefore direct and internally consistent. 
Straight number counts in the quasar and
background fields show a clear excess 
of galaxies at faint magnitudes (see Wold et al.\ 2000a; 2000b), and we also 
find tentative evidence for clusters at the quasar redshifts in the form of a 
red sequence present in the richest RQQ fields, see Fig.\ref{figure:fig3}.

\section*{A link between radio luminosity and environmental richness?}

This section treats the environment of the
RLQs.
As discussed in the beginning, our aim with selecting sources from different
surveys with different flux-density limits was to overcome the
luminosity-redshift degeneracy.

\begin{figure}[h]
\includegraphics{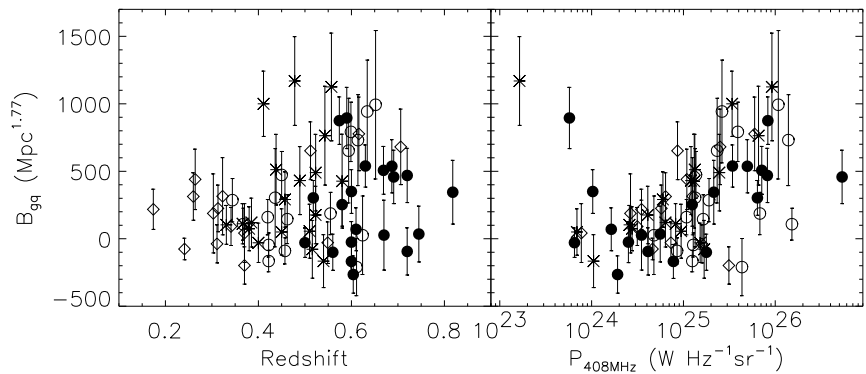}
\caption{Clustering amplitudes in RLQ fields as a function of redshift (left plot)
and radio luminosity at 408 MHz (right plot). 
Filled circles: this work, asterisks: Ellingson et al.\ (1991), open circles:
Yee \& Green (1987), diamonds: Yee \& Ellingson (1993).}
\label{figure:fig4}
\end{figure}

In our data, we found
a hint of a correlation between the clustering amplitude and the
radio luminosity of the quasars. To further investigate this,
we added 51 steep-spectrum quasars from the work by Yee \& Green (1987),
Ellingson et al.\ (1991) and Yee \& Ellingson (1993), mostly at
redshifts $0.2 < z <0.6$. These data are plotted in Fig.~\ref{figure:fig4},
where it can be seen that the clustering amplitude correlates
more with radio luminosity than with redshift.

To analyze this, we used Spearman's partial rank correlation 
coefficients giving the correlation coefficient between two variables holding
the third constant. The correlation coefficient between 
$B_{\rm gq}$ and $z$, holding $L_{408}$ constant, is 0.10 with a
0.8$\sigma$ significance, whereas the correlation coefficient between
$B_{\rm gq}$ and $L_{408}$, holding $z$ constant, 
is 0.4 with a 3.4$\sigma$ significance. We thus find no evidence for
the cosmic evolution in $B_{gq}$ as has been claimed for RLQs
and radio galaxies (Yee \& Green 1987; Ellingson et al.\ 1991; Hill \& Lilly
1991). 

The correlation between $B_{gq}$ and $L_{408}$ is weak with much scatter, 
nevertheless, it is significant. Does this imply that 
the large-scale radio emission in the RLQs is 
affected by the environment? Models of radio sources
certainly suggest this, where the minimum energy density in the radio lobes scales 
with the ram pressure at the working surface, implying that the synchrotron 
luminosity scales with the external gas density \cite{mrs93}.
Wold et al.\ (2000a) 
compare the data with a model which assumed {\it all} the variation 
in $L_{408}$ is due to the differences in the environments, but find that the 
predicted correlation is much steeper than observed. It 
is thus possible that both the environmental density and the bulk kinetic
power in the radio jets determine the radio luminosity. 
Alternatively, the relation between $B_{\rm gq}$ and 
$L_{408}$ may
just reflect an increasing mass of the host. This is 
possible if the radio luminosity is determined mostly 
by black hole mass, and if galaxies with massive black 
holes prefer rich environments.

\section*{Summary}

\begin{enumerate}
\item Both the radio-loud and the radio-quiet quasars 
studied in this survey live in a diversity of 
environments, from field-like environments to what appears to be rich 
galaxy clusters. Only about 10 per cent of the quasars live in 
relatively rich clusters of Abell richness class 1--2, and 
approximately 45 per cent live in field-like environments. 
\item The average environmental richness in the RLQ and the RQQ
fields is statistically indistinguishable, corresponding to  
groups or poorer clusters. We therefore find that on scales of 0.5 Mpc
there is no difference in the environments of the RLQs and the
RQQs.
\item We find no evidence of an evolution with epoch in the environments
of RLQs. Instead, the claimed evolution with redshift
might have been caused by selection effects in flux-limited samples. 
The true underlying correlation may be that of environmental richness 
with radio luminosity.
  
\end{enumerate}


\end{document}